\documentclass[11pt]{article}

\pdfoutput=1
\usepackage{amsthm, amssymb,latexsym,amsfonts}
\usepackage{putex}

\usepackage{comment}
\usepackage{graphicx}
\usepackage{caption}
\usepackage{amsmath}
\usepackage[usenames,dvipsnames,table]{xcolor}
\usepackage{array}
\usepackage{subcaption}
\usepackage{epstopdf}
\usepackage{enumerate}
\usepackage{cite}
\usepackage{tensor}
\usepackage{slashed}
\usepackage[export]{adjustbox}
\usepackage[aligntableaux=center]{ytableau}
\usepackage{dsfont}
\usepackage[utf8]{inputenc}
\usepackage[
      colorlinks=true,
      linkcolor=blue,
      urlcolor=blue,
      filecolor=black,
      citecolor=red,
      ]{hyperref}
\newcommand{\RNum}[1]{\uppercase\expandafter{\romannumeral #1\relax}}

 \usepackage[parsep]{collref}

\newcommand {\be} {\begin {equation}}
\newcommand {\ee} {\end {equation}}

\newcommand {\bes} {\begin {equation*}}
\newcommand {\ees} {\end {equation*}}

\newcommand{\CP}{\mathbb{CP}}

\newcommand{\beq}{\begin{equation}}
\newcommand{\eeq}{\end{equation}}

\newcommand{\ov}{\over}

\def\ie{\begin{equation}\begin{aligned}}
\def\fe{\end{aligned}\end{equation}}

\newcommand{\la}{\langle}

\newcommand{\B}{{\beta}}

\numberwithin{equation}{section}

\def\<{\langle}
\def\>{\rangle}

\newcommand{\foot}{\footnote}
\newcommand{\ci}{\cite}
\def\ov{\over}
\newcommand{\rf}[1]{(\ref{#1})}
\def\no{\nonumber}

\def \adss {$\text{AdS}_{5}\times S^{5}$\ }

\def \ha {{1 \ov 2}}

\def \l {\lambda} 
\def \no {\notag}
\def \iffa {\iffalse} 
\def \ed {\end{document}}
\def \F {{_F}}   \def \B {{_B}}

\def \ZZ  {{\mathbb Z}}

\def \la {\label}
\def \lpl {{\ell_{\rm pl}}} 
\def \l {\lambda} 
\def \gs  {g_{\rm s}}  \def \CP  {{\rm CP}}

\def \str {{\rm s}}
\def \cc {{\rm c}} 
\def \ha {{1\ov 2}}
\def \AA {{\rm A}}
\def \G  {\Gamma} 

\begin{document}

\preprint{PUPT-2639\\ Imperial-TP-AT-2023-02}
\institution{PU}{ $^{a}$ Department of Physics, Princeton University, Princeton, NJ 08544, USA }
\institution{}{ $^b$ Blackett Laboratory, Imperial College, London SW7 2AZ, UK}
\title{Wilson loops at large $N$ and the quantum M2 brane} 

\authors{Simone Giombi$^a$ and Arkady A. Tseytlin$^{b,}$\footnote{ Also   on leave from 
the  Inst. for Theoretical and Mathematical Physics (ITMP)  and Lebedev Inst.}}

\abstract{
The Wilson loop operator in the $U(N)_k \times U(N)_{-k}$ ABJM theory at  large $N$    and fixed level $k$ has a dual description in terms of a wrapped M2 brane in the M-theory background AdS$_4 \times S^7/\mathbb Z_k$. We consider the localization result for the $1\ov 2$-BPS circular Wilson loop expectation value $W$ in this regime, and compare it to the prediction of the M2 brane theory. The leading large $N$ exponential factor 
is matched  as expected  by  the classical action of the M2 brane solution with AdS$_2\times S^1$ geometry. We show that the  subleading   
$k$-dependent prefactor in $W$ is also exactly reproduced 
  by the one-loop term in the partition function of the wrapped M2 brane (with all Kaluza-Klein modes included). This appears to be the first case of an exact matching of the overall numerical prefactor in the Wilson loop expectation value against the dual holographic result. It provides an example of a consistent quantum M2 brane computation, suggesting various generalizations.}

\maketitle

\tableofcontents

\section{Introduction}

The existence of a consistent quantum  supermembrane (or M2 brane) theory  remains  an enigma (see e.g. \ci{Duff:1996zn,Nicolai:1998ic}).
The corresponding  3d world-volume theory is formally non-renormalizable, apparently requiring  
a specific definition like a built-in cutoff. Nevertheless, some simple semiclassical computations to one-loop order can
still be done in a  straightforward way, 
as one-loop  corrections in 3d field theory are free of logarithmic UV divergences,  
see  e.g. \cite{Duff:1987cs,Bergshoeff:1987qx,Mezincescu:1987kj,Forste:1999yj}  or  more recent work in \ci{Drukker:2020swu}.

 In this paper, we present another example of one-loop calculation in the M2 brane theory, which provides further evidence that the quantization of the supermembrane might be under good control, at least within the semiclassical expansion. 

The   AdS$_4$/CFT$_3$  duality between  the $U(N)_k\times U(N)_{-k}$  ABJM theory \cite{Aharony:2008ug} 
  and M-theory  on AdS$_4\times S^7/\ZZ_k$  provides  a remarkable 
  opportunity  to shed light  on the  properties of 
  the  quantum M2 brane theory  by testing its  
  predictions  against  exact  results  in 3d  superconformal gauge theory. 
In the large $N$ limit with $k$ fixed, the holographic dual of a Wilson loop in the fundamental representation is expected to be an  M2 brane wrapping the M-theory circle direction. Note that this limit is different from the standard large $N$ 't Hooft limit, where $N$ and $k$  
are taken  to  be large  with $\l=N/k$ fixed, and in which Wilson loops are described by fundamental strings in type IIA string theory  in AdS$_4\times \CP^3$. 

For fixed $k$, 
the large  $N$  expansion of the Wilson loop operator in the ABJM theory corresponds to the expansion in the large effective M2 brane tension $R^3 T_2\sim \sqrt{ N k } $,  where $R$ is the curvature radius  of AdS$_4\times S^7/\ZZ_k$   and 
$T_2 = {1\ov   (2 \pi)^2\ell_{\rm pl}^3}$.  

Our starting point will be an analytic expression for the expectation value of the $1\ov 2$-BPS circular Wilson loop
  in the  ABJM theory  
  derived  using supersymmetric localization in \cite{Klemm:2012ii} 
  (see  also \cite{Drukker:2008zx,Kapustin:2009kz,Drukker:2009hy,Drukker:2010nc,Herzog:2010hf,Fuji:2011km,Marino:2011eh})\foot{We use the same  normalization  of $\langle W_{\frac{1}{2}}\rangle$ as 
in \ci{Giombi:2020mhz}.}  
\begin{equation}
\langle W_{\frac{1}{2}}\rangle = \frac{1}{2\, \sin(\frac{2\pi}{k})} \frac{{\rm Ai}\left[C^{-\frac{1}{3}}\left(N-\frac{k}{24}-\frac{7}{3k}\right)\right]}{{\rm Ai}\left[C^{-\frac{1}{3}}\left(N-\frac{k}{24}-\frac{1}{3k}\right)\right]}\ . 
\label{W-Airy}
\end{equation}
where ${\rm Ai}(z)$ is the Airy function, and $C=2/(\pi^2 k)$. 
This expression resums all of the perturbative $1/N$ corrections at fixed $k$.\foot{The leading behavior of the Wilson loop in this large $N$, fixed $k$ limit was first discussed in \cite{Herzog:2010hf}, but the $k$-dependent sine prefactor, which comes from the fluctuations about the large $N$ saddle point, was not derived there.}  

In  order to compare to the semiclassical expansion in the M2 brane world-volume theory, 
one is to expand (\ref{W-Airy}) at  large $N$  with  fixed  $k$,  which gives    
\begin{equation}
\langle W_{\frac{1}{2}}\rangle = \frac{1}{2\sin(\frac{2\pi}{k})}e^{\pi \sqrt{\frac{2N}{k}}}\left[
1-\frac{\pi  \left(k^2+32\right)}{24 \sqrt{2} \, k^{3/2}}\frac{1}{\sqrt{N}}+O(\frac{1}{N})
\right] \ . 
\label{loc-largeN}
\end{equation}
As we will show in Section \ref{M2-Z} below, the exponential factor in \rf{loc-largeN}  is reproduced by  the classical action of the 
M2 brane  with AdS$_2\times S^1$ world-volume, 
while the $k$-dependent  prefactor $(2 \sin\frac{2\pi}{k})^{-1}$ 
is 
 matched precisely  by  the one-loop correction coming from the functional determinants of the quantum fluctuations around this  M2 brane solution. 
  
Higher-order  $1/(\sqrt N)^n$  terms  in \rf{loc-largeN} 
are expected to
represent
  higher-loop corrections in the semiclassical expansion of the partition function of the quantum M2 brane theory,   and 
   checking this  is  a very interesting but challenging future problem. 
  We comment on  this   and   other generalizations  in Section \ref{Concl}.

\section{The AdS$_2\times S^1$ M2 brane in AdS$_4\times S^7/\ZZ_k$} 
\label{M2-Z}

The  AdS$_4\times S^7/\ZZ_k$ metric is given by\foot{Explicitly,  in \rf{22} one has 
$ ds^2_{{\rm CP}^3}=  {d w^s d \bar w^s\ov 1 + |w|^2}   - { w_r \bar w_s   \ov (1 + |w|^2)^2 } d  w^s d   \bar w^r $
  and $dA=   i [{ \delta_{sr} \ov 1 + |w|^2} - {w_s \bar w_r\ov (1 + |w|^2)^2} ] d w^r \wedge d \bar w^s   $  ($s,r=1,2,3)$.  }
\begin{align}
ds^2 =& \frac{R^2}{4}ds^2_{{\rm AdS}_4} + R^2 ds^2_{S^7/\ZZ_k} \ , 
\label{AdS4S7}\\
ds^2_{{\rm AdS}_4} = &\frac{1}{z^2}\left(-dt^2+dz^2  +dx_1^2+dx_2^2\right)\,, 
\label{AdS4}
\\
\la{22}
ds^2_{S^7/\ZZ_k} =& ds^2_{{\rm CP}^3}+\frac{1}{k^2}\left(d\varphi+k A\right)^2\ , \ \ \ \ \ \ \ \qquad  \varphi \equiv  \varphi+2\pi\ .
\end{align}
The 11d supergravity background also includes the 4-form field strength
\begin{equation}
F_4 = dC_3 = -\frac{3}{8}\frac{R^3}{z^4}dt\wedge dx_1\wedge dx_2\wedge  dz\,.  
\end{equation}
The radius $R$ in units of the 11d  Planck  length $\lpl$ is related to the parameters $N$ and $k$ of the dual ABJM gauge theory 
by\foot{We are ignoring  higher order  corrections 
   to this relation (due to $N\to N - {1\ov 24} (k - { k^{-1}}) $  \ci{Bergman:2009zh})
as they will not be relevant  for  the leading   large $N$    M2 brane contribution considered below.} 
\begin{equation}
\left(\frac{R}{\ell_{\rm pl}}\right)^6 = 2^5 \pi^2 N k\,.
\label{R-map}
\end{equation}
The world-volume action for a probe  M2 brane   in this    background  is given by  \ci{Bergshoeff:1987cm,deWit:1998yu}\foot{The AdS$_4\times S^7$  background 
is a  limit of the 11d   supergravity solution describing $N$ M2 branes \ci{Duff:1990xz}. 
Here  we consider a probe  M2 brane  intersecting  the boundary (or  coincident M2 branes) 
over a line, which is a BPS configuration   like  a fundamental string  intersecting D3 branes or D2 branes 
in the  analogous   \adss  and AdS$_4 \times {\rm CP}^3 $  cases.}
\begin{equation}
S_{\rm M2} = T_2 \int d^3\sigma \sqrt{-{\rm det}\, g} + T_2 \int  C_3 \   + \ {\rm fermionic \ terms}\ , 
\label{M2-action}
\end{equation}
where  the M2 brane tension 
 is\foot{The  universal expression  for the M2 brane tension  in terms of  the coefficient $\kappa_{11}$ in the 11d  theory action ${1\ov 2 \kappa_{11}^2} \int d^{11} x \sqrt{ -g}\,  R + ...$  is \ci{Klebanov:1996un}:
$T_2 = ( 2\pi)^{2/3} (2 \kappa_{11}^2)^{-1/3}$.  Here $\ell_{\rm pl}$ is defined so that 
$2 \kappa_{11}^2 = (2 \pi)^8 \ell_{\rm pl}^9$  as, e.g., in \ci{Bagger:2012jb}.}
\begin{equation}
T_2 = \frac{1}{(2\pi)^2} \frac{1}{\ell_{\rm pl}^3}\,.
\label{T2-map}
\end{equation}

\subsection{Classical M2 brane solution}

The action \rf{M2-action} admits a simple classical solution
 given by the M2 brane
 wrapping the M-theory circle direction (the $\varphi$ angle in (\ref{AdS4S7})), and occupying the AdS$_2$ subspace of AdS$_4$  spanned by the coordinates $t$, $z$ in (\ref{AdS4}).
  The resulting membrane has the  AdS$_2\times S^1$ world-volume  geometry 
  and is dual to the $1\ov 2$-BPS Wilson loop along the $t$ direction at the boundary of AdS$_4$.
    By an appropriate Wick rotation and coordinate transformation, one may obtain in the same way the solution dual to the circular Wilson loop, for which the AdS$_2$ factor is just the Euclidean hyperbolic disk with 
    circular boundary. 

The value of the classical action \rf{M2-action}  for this AdS$_2\times S^1$ solution is simply given by\footnote{Note 
that there is no contribution from the Wess-Zumino term involving the 3-form  field in \rf{M2-action}.}
\begin{equation}
S^{\rm cl.}_{\rm M2} = T_2 R^3 \frac{1}{4}{\rm vol}({\rm AdS}_2) \frac{2\pi}{k} \ , 
\label{SM2-action}
\end{equation} 
where  $1\ov 4$ comes from the  AdS$_4$ radius in (\ref{AdS4S7}), and the
 $2\pi/k$  is the length of the M-theory $\varphi$-circle in \rf{22}. 
 Using (\ref{R-map}) and (\ref{T2-map}), and the well-known value of the regularized volume of the unit-radius
  hyperbolic disk, ${\rm vol}({\rm AdS}_2)=-2\pi$, this gives
    (we  always assume $k >0$)
\begin{equation}\la{29}
S_{\rm M2}^{\rm cl.} = -\pi \sqrt{\frac{2N}{k}}\,.
\end{equation}
 Thus 
 $e^{-S_{\rm M2}^{\rm cl.}}$ precisely matches the exponential in the localization prediction (\ref{loc-largeN}).\foot{This AdS$_2\times S^1$ solution was  considered
    in \ci{Sakaguchi:2010dg}  but the calculation of the regularized classical action and its connection to the Wilson loop expectation value was not discussed there.
    Various wrapped  BPS M2 brane solutions  in AdS$_4 \times X^7$ and  matching of their classical actions to exponents in the 
    localization prediction
    for the corresponding 3d  BPS Wilson loops, including the present $1\ov2$-BPS one in   AdS$_4 \times S^7/\ZZ_k$, 
    was demonstrated 
     in  \ci{Farquet:2013cwa}. Supersymmetries of similar M2-brane solutions  were considered also in 
  \ci{Lietti:2017gtc}).}  
 
  In the next section we will  also compute the one-loop correction  to the M2 brane partition function 
  due to the quantum fluctuations about this  classical solution,  and  will 
   reproduce precisely  the prefactor in (\ref{loc-largeN}).

Let us note that  in the case of the $1\ov 2$-BPS Wilson loop along the infinite straight line, 
one should  use in (\ref{SM2-action}) the regularized volume of  AdS$_2$ in Poincare coordinates
which is zero,  thus getting    $S_{\rm M2}^{\rm cl.} =0$,  
 consistently with $\langle W_{1\ov 2}\rangle = 1$ in this case. 
 All  quantum corrections also vanish
  here   since  the AdS$_2$ space  is homogeneous and
   hence the quantum  M2 brane free energy is proportional to ${\rm vol}({\rm AdS}_2)$ to all orders.\foot{This is the  
   same argument  that applies  also   in  the similar  \adss   string context  \ci{Drukker:2000ep,Buchbinder:2014nia}.}

\subsection{One-loop correction}

Starting  with the action (\ref{M2-action}) one may expand it near  a  classical solution to quadratic order  fixing a 3d reparametrization and  $\kappa$-symmetry gauge  to get an action for  8+8 physical  3d  fluctuation fields. 
The resulting spectrum 
 of the quantum fluctuations around the  above AdS$_2\times S^1$ solution was obtained in ref. \cite{Sakaguchi:2010dg}, that we follow below.

It is natural to chose  a static gauge  identifying   two  membrane coordinates $\sigma_{1},\sigma_2$ in \rf{M2-action}
with the  AdS$_2$ directions and the third $\sigma_3$ with the $S^1$ angle $\varphi$. 
 After a Kaluza-Klein (Fourier) expansion  of the  3d fields in the periodic   coordinate $\sigma_3 $,
 one obtains a tower of bosonic and fermionic fluctuations that can be viewed as
 2d fields propagating on  the (unit-radius) AdS$_2$ background. Thus one gets  an equivalent 2d theory with an  infinite number of fields.   
 
  The bosonic fluctuations in the two transverse directions within AdS$_4$  give 
   a tower of complex scalar fields $\eta_n$ (two real scalars for each $n$) with masses
\begin{equation}
m_{\eta_n}^2 = \frac{1}{4}(kn-2)(kn-4)\,,\qquad n \in \mathbb{Z}\,,
\end{equation}
while from the fluctuations in the  six  ${\rm CP}^3$ directions one finds a tower of 3 complex fields $\zeta^s_n$ ($s=1,2,3$) with masses
\begin{equation}
m_{\zeta^s_n}^2 = \frac{1}{4}kn(kn+2)\,,\qquad n\in \mathbb{Z}\,.
\end{equation}
For the fermionic fluctuations, the KK reduction leads to a tower of eight two-component spinors $\vartheta^A_n$ ($A=1,\ldots, 8$) for each value of the KK mode number $n$, with masses given by\footnote{In more detail, in the notations of \cite{Sakaguchi:2010dg} the eight fermions are labelled by $\vartheta^{\alpha_1\alpha_2\alpha_3\alpha_4}$ where $\alpha_i=\pm 1$ and $\alpha_1\alpha_2\alpha_3\alpha_4=-1$. The masses of the KK fermion tower are then given by $m_{\vartheta_n} = \frac{1}{4}(3\alpha_1+\alpha_2+\alpha_3+\alpha_4-2kn)$, $n\in\mathbb Z$, which leads to the spectrum summarized in (\ref{fermi-mass}).  Also, the  resulting 
action  for  the  supermembrane  fermions  that are 
space-time spinors  may be reinterpreted as an action  for  
2d  Majorana   spinors.} 
\begin{equation}
m_{\vartheta^a_n} = \frac{kn}{2}\pm 1 ~~~(\mbox{3+3 modes})\,,\qquad \qquad
m_{\vartheta^i_n} = \frac{kn}{2} ~~~(\mbox{2 modes})\,,\qquad n\in \mathbb{Z}\,.
\label{fermi-mass}
\end{equation}
The  above masses   explicitly depend on  the integer $k$   which  is the inverse radius of the  $\varphi$-circle in \rf{22}. 
Thus,  in the type IIA string
 limit  $k\to \infty$, all  KK modes  with $n\not=0$  become   infinitely heavy. 

For $n=0$, this spectrum coincides (as    
expected upon double dimensional reduction  \ci{Duff:1987bx})  with the spectrum of bosonic and fermionic fluctuations around the  corresponding AdS$_2$ string solution  in the type 
IIA superstring  theory on AdS$_4 \times {\rm CP}^3$ 
   \cite{Kim:2012tu,Giombi:2020mhz}: we get 2  scalars of $m^2=2$,  6  scalars of $m^2=0$ scalars, 
 3+3 fermions  of $m=\pm 1$  and 2 fermions  of  $m=0$.

One can also check that the  full   spectrum is consistent with 2d supersymmetry. 
The bosonic and fermionic masses in a ${\cal N}=1$ supermultiplet in AdS$_2$   containing one real scalar and a Majorana fermion   are related as  (see e.g.  \ci{Sakai:1984vm})
\begin{equation}\la{213}
m_{\B}^2 = m_{\F} (m_{\F}-1) \ . 
\end{equation}
Indeed,  the bosonic and fermionic  modes  listed above  can be grouped so that 
their  masses   satisfy this  relation. 

A stronger consistency test of the spectrum is obtained by checking the vanishing of the vacuum energy in
 Lorentzian AdS$_2$
 in global coordinates
(as that happens also in the simple  case of  the flat toroidal  M2  brane \ci{Duff:1987cs,Gandhi:1988sh}). 
 The vacuum energies for massive bosons and fermions in AdS$_2$ are given by (see e.g. \cite{Drukker:2000ep})
\begin{equation}\la{214}
E_\B (m_\B) = -\frac{1}{4}\big(m^2_\B+  \frac{1}{6}\big)\,,\qquad \ \ \ 
E_\F(m_\F) = \frac{1}{4}\big(m^2_\F -\frac{1}{12}\big)\,.
\end{equation} 
We   find that  the total vacuum energy 
in  the present  case  is  zero  separately  for  each  KK   level $n$ 
\begin{align}
&\qquad \qquad \qquad\qquad \qquad \qquad E^{\rm tot}=\sum_{n=-\infty}^\infty  E^{\rm tot}_n\ , \la{215} \\
&E^{\rm tot}_n = -\frac{1}{4}\left[ \frac{2}{4}(kn-2)(kn-4)
+\frac{6}{4}kn(kn+2)  -3(\frac{kn}{2}+1)^2-3(\frac{kn}{2}-1)^2-2(\frac{kn}{2})^2
+2\right] = 0\ . \la{255}
\end{align}

Using the above spectrum, we can derive the one-loop correction to the partition function of the M2 brane
theory expanded  around the Euclidean AdS$_2\times S^1$  solution with circular boundary. The semiclassical partition function is given by
\begin{equation}
Z_{\rm M2} = Z_{1} e^{-S_{\rm M2}^{\rm cl.}}\left[1+O\Big(\frac{1}{R^3T_2}\Big)\right] \ ,  \la{216}
\end{equation}
where the one-loop term $Z_1$ is the ratio of the determinants of the corresponding fluctuation operators
\begin{equation}
Z_{1} =\prod_{n\in \mathbb{Z}} \frac{\big[{\rm det}(-\nabla^2+\frac{R^{(2)}}{4}+(\frac{kn}{2}+1)^2)\big]^{\frac{3}{2}}
\,
\big[{\rm det}(-\nabla^2+\frac{R^{(2)}}{4}+(\frac{kn}{2}-1)^2)\big]^{\frac{3}{2}}
\,
{\rm det}(-\nabla^2+\frac{R^{(2)}}{4}+(\frac{kn}{2})^2)
}{{\rm det}\big(-\nabla^2+\frac{1}{4}(kn-2)(kn-4)\big)\,
\big[{\rm det}\big(-\nabla^2+\frac{1}{4}kn(kn+2)\big) \big]^3}\,.\la{217}
\end{equation}
Here  $R^{(2)}=-2$ is the curvature of AdS$_2$.\footnote{As usual,  the form 
$- \nabla^2   + \frac{R^{(2)}}{4} + m^2$   of the second order differential operator
assumed to be  defined on 2d  Majorana   spinor fields 
 arises from squaring the Dirac operator (in our  conventions  its 
  determinant is counted as if it acts  on a single-component  real field). 
    }
The $n=0$   factor in \rf{217}
 is of course the same as  the one-loop partition function  \cite{Kim:2012tu,Giombi:2020mhz}  for  the  fluctuations near  the 
 corresponding type IIA   
 AdS$_2$ string worldsheet ending on a  circle at the boundary of AdS$_4 \times \rm CP^3$. 

The functional determinants in \rf{217} 
may be computed by the  standard  AdS$_d$ spectral zeta-function techniques (as  was done in  the similar AdS$_2$
 string case  in  e.g.  \ci{Drukker:2000ep,Buchbinder:2014nia,Giombi:2020mhz}).
 For a massive boson, one has 
\begin{equation}
\G_{1_{B} } =\ha \log{\rm det}(-\nabla^2+m_\B^2) = -\ha \zeta(0;m_\B^2)\log(\Lambda^2)-\ha \zeta'(0;m_\B^2) \ , \la{219}
\end{equation} 
where $\Lambda$ is a 2d UV cutoff, AdS$_2$  is assumed to have unit radius and 
\begin{align}
&\zeta_\B(0;m_\B^2) = \frac{m_\B^2}{2}+\frac{1}{6}\ , \la{200}\\
&\zeta_\B'(0;m_\B^2) = -\frac{1}{12}(1+\log 2)+\log \AA-\int_0^{m_\B^2+\frac{1}{4}} dx\ \psi(\sqrt{x}+\frac{1}{2})\ . \la{220}
\end{align}
Here  $\AA$ is the Glaisher's constant  and $\psi(x) = \Gamma'(x)/\Gamma(x)$.
Similarly, for  a  massive fermion 
\begin{align}\la{221}
\G_{1_{F} }= -\ha \log{\rm det}(-\nabla^2+\frac{R^{(2)}}{4}+m_\F^2) &= -\ha \zeta_\F(0;m_\F)\log(\Lambda^2)-\ha \zeta'_\F(0;m_\F) \ , \\ 
\zeta_\F(0;m_\F) &= -\frac{m_\F^2}{2}+\frac{1}{12}\ , \la{222}\\
\zeta'_\F(0;m_\F) & = -\frac{1}{6}+2\log \AA+|m_\F|+\int_0^{m_\F^2} dx\ \psi(\sqrt{x})\,.\la{223}
\end{align}

Using these expressions   we can first verify the 
cancellation of the logarithmically divergent part  of the one-loop free energy $\Gamma_1 = -\log Z_1$ in \rf{217}.
 Indeed, from the above calculation of the vacuum energy, one can 
 see that the sum over the bosonic and fermionic masses at each KK level 
   $n$ satisfies $\sum(m_\B^2-m_\F^2)=-2$. Then 
  the total coefficient of the logarithmic divergence  in the sum of the corresponding terms in  \rf{219},\rf{200}  and  \rf{221},\rf{222}  over the spectrum  is
\begin{equation}
\zeta_{\rm tot}(0) = {1\ov 2} \sum_{n\in \mathbb{Z}}  \big(-2 + 4\big) =     \sum_{n\in \mathbb{Z}} 1  = 1 +  2\zeta_R(0) = 0\,, \la{225}
\end{equation}
where we have used the Riemann zeta-function regularization to evaluate the (linearly divergent) sum.
Note that  the contribution of  all massive  KK  modes at  non-zero $n$  levels
cancels 1  coming from the   $n=0$  modes,
 i.e.  cancels the logarithmic UV divergence  that was present in the similar computation 
 in the AdS$_4 \times \rm CP^3$ superstring regime \ci{Giombi:2020mhz}.

The vanishing of the logarithmic divergence in the free energy  was actually   expected, as the M2 brane  theory we started with 
 is three-dimensional, and there are no logarithmic divergences in the  corresponding 
 functional determinants in 3d. The  reduction  to 2d with all KK modes included 
  cannot produce logarithmic divergences that  were not present in the 3d formulation.\foot{In general, 
  to match the divergences  in   3d and 2d formulations   one  is to  regularize  the 2d theory in a way 
  consistent with  symmetries of the 3d theory. 
  An  analytic regularization   like $\zeta$-function one that discards  power divergences  is sufficient  for this purpose 
   in the present context.}
  
The one-loop free energy is  thus   finite  and is  given by
\begin{align}
&\Gamma_1 =- \log Z_1  =  -\frac{1}{2}\zeta'_{\rm tot}(0)\ , \la{226} 
\end{align}
where according to \rf{217}
\begin{align}
\la{227}
&\zeta'_{\rm tot}(0) = \sum_{n\in \mathbb{Z}} \zeta'_{\rm tot}(0;n)\ , \\
&\zeta'_{\rm tot}(0;n) = 2 \zeta'_\B(0;\frac{1}{4}(kn-2)(kn-4))
+6 \zeta'_\B(0;\frac{1}{4}kn(kn+2))\no \\
&~~~~~~~~~~~~~~~~
+3 \zeta'_\F(0;\frac{kn}{2}+1)
+3 \zeta'_\F(0;\frac{kn}{2}-1)+2\zeta'_\F(0;\frac{kn}{2})\,.\la{228}
\end{align}
Summing up  the bosonic and fermionic contributions, some  remarkable simplifications occur. 
Combining  the contributions of the  positive and negative modes (so that below $n\ge 0$)
we find the following result\foot{The vanishing of the  $n=0$ contribution, i.e.  that  of the type IIA string modes, was  already observed   in \ci{Giombi:2020mhz}.} 
\begin{equation}\la{229}
\zeta'_{\rm tot}(0;n)+\zeta'_{\rm tot}(0;-n) = 
\begin{cases}
&-2 \log\left(\frac{k^2n^2}{4}-1\right)\,,\qquad kn>2\\
&\ \ \log(\pi^2)\,,\qquad \qquad \qquad \    kn=2\\
&-\log(\frac{9}{4})\,,\qquad\qquad\qquad  \ \   kn=1\\
& \ \ 0\,,\qquad \qquad\qquad \qquad   \ \ \ \  n=0
\end{cases}
\end{equation}
If we assume that $k>2$, only  the $n=0$ and $kn>2$ cases in \rf{229} occur, 
and  the complete one-loop free energy is given by the following simple result
\begin{equation}
\Gamma_1 = \sum_{n=1}^{\infty} \log\left(\frac{k^2n^2}{4}-1\right)
=2\sum_{n=1}^{\infty} \log(\frac{kn}{2})+ \sum_{n=1}^{\infty}\log\left(1-\frac{4}{k^2n^2}\right)\,.\la{230}
\end{equation}
Using again the  Riemann zeta-function regularization   
($ \zeta_R(0)= - {1\ov 2}, \ \zeta'_R(0)= - {1\ov 2} \log (2 \pi)$) we get
\begin{equation}
2\sum_{n=1}^{\infty} \log(\frac{kn}{2}) = 2 \zeta_R(0)\log(\frac{k}{2}) 
-2 \zeta'_R(0) = -\log(\frac{k}{4\pi})\,. \la{231}
\end{equation}
The second sum in \rf{230}  is finite and given by
\begin{equation}
\sum_{n=1}^{\infty}\log\Big(1-\frac{4}{k^2n^2}\Big) 
= \log\Big[\prod_{n=1}^{\infty}\big(1-\frac{4}{k^2n^2}\big)\Big] 
=\log \Big[\frac{k}{2\pi}\sin\big(\frac{2\pi}{k}\big) \Big]\,. \la{232}
\end{equation}
Here we used the  Euler's  expression  for the sine  as a product of its zeros,  \ \ 
 $\sin(\pi x)=\pi x \prod_{n=1}^{\infty}(1-\frac{x^2}{n^2})$. 
 
 Combining \rf{231} and \rf{232}  we get 
 the final result for the one-loop partition function for $k>2$ 
\begin{equation}
Z_1 = e^{-\Gamma_1} = \frac{1}{2\sin(\frac{2\pi}{k})}\,, \la{233}
\end{equation} 
which is thus  in precise agreement with the localization result in \rf{loc-largeN}.

Let us now  discuss the  special  cases of  $k=1,2$ 
 which require a separate treatment.\footnote{Recall that for these values of $k$, the supersymmetry of the M-theory background (\ref{AdS4S7}) and of the dual gauge theory is enhanced from
  ${\cal N}=6$ to ${\cal N}=8$ \ci{Halyo:1998pn,Aharony:2008ug,Bagger:2012jb,Gustavsson:2009pm}.} 
 For $k=1$, all of the cases listed in (\ref{229}) occur in the sum over the KK modes,  i.e. 
\begin{equation}\la{234}
\Gamma_1^{k=1} = \frac{1}{2}\log(\frac{9}{4})-\frac{1}{2}\log(\pi^2)+\sum_{n=3}^{\infty}\log\Big(\frac{n^2}{4}-1\Big)\,.
\end{equation} 
The infinite sum here can be evaluated   in a similar way  (i.e. using $\zeta$-function regularization) 
 as  explained above for \rf{230} 
\begin{equation}
\sum_{n=3}^{\infty}\log\Big(\frac{n^2}{4}-1\Big) = 2\sum_{n=3}^{\infty}\log(\frac{n}{2})+\sum_{n=3}^{\infty}\log(1-\frac{4}{n^2})=\log(16\pi)-\log 6\,.\la{235}
\end{equation}
Then from \rf{234} we get 
$\Gamma_1^{k=1}=\log 4$  and thus
\begin{equation}
Z_{1}^{k=1} = \frac{1}{4}\,.\la{236}
\end{equation}
Similarly, for $k=2$ we find 
\begin{align}
\Gamma_1^{k=2} = -\frac{1}{2}\log(\pi^2)+ &\sum_{n=2}^{\infty}\log\left(n^2-1\right) =0\,,\la{237}
\\
\ \ \ \ \ Z_{1}^{k=2} & =1\,.\la{238}
\end{align}
These results can not be  directly  compared to localization,  as  the  result  (\ref{W-Airy})  of \cite{Klemm:2012ii}  
 is  singular   for  $k=1,2$.\foot{Remarks
    that localization  result is  divergent for $k=1,2$ appeared, e.g.,  in  \ci{Klemm:2012ii,Honda:2013nmk,Hirano:2014bia,Hatsuda:2016rmv}.} 
    It might be  that the  derivation of  (\ref{W-Airy})  
   in \cite{Klemm:2012ii}  is to be reconsidered  specifically  for   $k=1,2$. 
   The matching in these  special  cases  thus   remains an open problem. 

\section{Concluding remarks}
\label{Concl}

Extending the above computation to higher loops in the semiclassical expansion
of the   partition function \rf{216} 
would allow one to compare the  quantum  M2 brane   prediction with the subleading terms in the expansion of the localization expression at large $N$. For instance, the term of order $1/\sqrt{N}$ in \rf{loc-largeN} should come from a 2-loop calculation in the M2 brane theory (recall that ${1\ov R^3 T_2} \sim {1\ov \sqrt N}$).  Apart from technical difficulties,  one issue  with  this  computation is  whether  the  2-loop correction will be  UV finite.

The  cancellation of  logarithmic divergences  (despite apparent non-renormalizability) 
 may happen   due  to the large amount of supersymmetry of the  supermembrane theory (cf.  \ci{Paccanoni:1989hd}).
 An example of a   cancellation  of 2-loop  UV divergences in a  formally non-renormalizable  theory  is 
 provided by the successful 
 computation of the  subleading  $ 1\ov \sqrt \l$  correction to the cusp anomalous   dimension
 $f(\l)= a_0 \sqrt \l + a_1 + {a_2\ov \sqrt \l} + ...$  in the  \adss   superstring theory 
 \ci{Roiban:2007jf,Roiban:2007dq,Giombi:2009gd},   that 
  matched the  corresponding term in the strong-coupling expansion of  $f(\l)$
 derived on the ${\cal N}=4 $ SYM side using integrability \ci{Basso:2007wd}
  (the analogous  2-loop computation in 
 the case of the AdS$_4 \times \CP^3$  string was done in \ci{Bianchi:2014ada}). 
 
An alternative possibility could  be that the M2  brane theory   has a built-in UV cutoff $ \Lambda \sim \lpl^{-1}  \sim T_2^{-1/3}$.
 However, then a logarithmically divergent term would  scale as $\log ( R \Lambda)= { 1\ov 6} \log (N k) + ...$ (see \rf{R-map}),
 but there is no  such  $\log N $ term in the localization expansion of the Wilson loop in \rf{loc-largeN}.\foot{The $\log N$ term is  known to be present in the free energy of the ABJM   theory on $S^3$  and  it was matched to the one-loop 11d supergravity result in \cite{Bhattacharyya:2012ye}, where it comes only from the zero-mode contribution.} This suggests that the logarithmic divergences may cancel at higher loops in the M2 brane theory, at least for such a $1\ov 2$-BPS observable. 


Let us now  comment on the 10d type IIA  string theory limit, which correspond to $k$ and $N$ both taken to be large, with the 't Hooft coupling $\l=N/k$ kept fixed. In this regime, the 11d  background \rf{AdS4S7}  
reduces to   AdS$_4 \times \CP^3$, and a Wilson loop operator is dual to an open string ending on a  loop at the boundary of AdS$_4$.  The  corresponding 
type IIA string coupling constant $ \gs $ and  the effective  string  tension 
$T={1\ov 2 \pi}  {R^2_{\rm s}\ov \ell_{\rm s}^2} $  
 are then  given by  \ci{Aharony:2008ug}
\be 
 \gs  = {   \sqrt \pi\,(2 \l)^{5/4}\ov  N}  \ , \qquad 
     \ \ \ \ \   T = {\sqrt{2\l }\ov 2}  \ , \qquad \ \ \  \l = {N \ov k}  \ .  \la{31}
\ee
Note that to be in the $\gs \ll  1$ and $\l \gg 1 $  regime, we  need  to assume that $ k \ll N \ll  k^5$. 

As was  pointed out in \ci{Giombi:2020mhz},   the  string partition function  computed 
near the AdS$_2$ minimal surface   representing the $\ha$-BPS circular Wilson loop in both  type IIB \adss 
and type IIA AdS$_4 \times \CP^3$ theories has  an expansion  in small $\gs$ and then 
 in large  tension $T$  of   the following   universal form\foot{The fact that string loop  corrections go as   powers of ${g^2_\str \ov T }$  not  just  of $g^2_\str$ (and thus are suppressed  by factors of  $1/T$)  should  be a consequence of the underlying supersymmetry of the theory.}
\begin{align} 
\langle  W_{1\ov 2} \rangle
= &\  e^{2\pi\,T}\,  {\sqrt T \ov \gs } \Big\{ \cc_{0} \big[ 1 + O (T^{-1}) \big] 
  + \cc_{1}  { g^2_\str \ov T } \big[ 1 + O (T^{-1}) \big]
+  \cc_{2}    \Big({g^2_\str \ov T } \Big)^2 \big[ 1 + O (T^{-1}) \big] + ... \Big\} \ .  \la{32}
\end{align}
In our present case, we can see that this is consistent with the  structure of the corresponding large $N$, large $k$ expansion of \rf{W-Airy},  according to which 
one  should get 
$\cc_{0} = {1\ov \sqrt{2\pi}}$, $\cc_{1} = {\pi \ov 12} \cc_{0} , $ etc. 
The  presence  of the overall $\sqrt T$ factor was shown in 
 \ci{Giombi:2020mhz}   to originate  from the 
 the leading 
 one-loop   string sigma model correction  on the disk   (it  is     related   to the  $n=0$    contribution in \rf{225}). 
 The  precise value of the one-loop coefficient $\cc_{0} = {1\ov \sqrt{2\pi}}$   was not so far derived directly 
 on  the string  side (that appears to  require a careful  normalization of the   measure in  the  superstring  path integral). 
Remarkably,  the  M2  brane one-loop computation described above  effectively determines this coefficient and,  moreover,  
the coefficients   of all of the leading large tension terms at higher genus (disk with handles).  
 Indeed,  comparing  \rf{32} to the corresponding  large $N$,  large $k$  
   expansion  of \rf{W-Airy}, 
      the leading large tension terms in 
  \rf{32} can be seen  \ci{Beccaria:2020ykg}  to arise from a resummed expression
  \be  \la{33}
  \langle  W_{1\ov 2} \rangle = {1  \ov 2  \sin \big( \sqrt\frac{\pi}{2}\,\frac{\gs}{\sqrt T}\big)   } \, e^{2\pi\,T}  \Big[1+ O(T^{-1})\Big] \ , \ee
   where    $\sqrt\frac{\pi}{2}\,\frac{\gs}{\sqrt T}= 2\pi\,\frac{\lambda}{N}= {2 \pi \ov k }$. Here 
    the sine factor is just the same as the one found in (\ref{loc-largeN}),(\ref{233})  (the exponential factor is also the same as in \rf{216},\rf{29}). Thus, the one-loop M2 brane correction  happens  to describe the leading  large tension  terms
   at all orders in the genus expansion in the type IIA string theory!\foot{This
    is somewhat analogous  to how a semiclassical   AdS$_2 \times S^2$ D3-brane solution  happens to capture the 
    leading resummed contribution  \ci{Drukker:2000rr} in the  $1\ov 2$-BPS circular Wilson  loop expectation value $\langle W_\ha \rangle =  { 1 \ov 2 \pi} \frac{\sqrt T}{\gs}\, e^{2\pi\,T \  +\    \frac{\pi}{12}\frac{\gs^{2}}{T} }\,\Big[1+O(T^{-1})\Big]$ in the ${\cal N}=4$ SYM or   \adss  string  case \ci{Drukker:2005kx}.}

\iffa
At the same time,  higher order $1/T$  terms  (or $\alpha'/R^2$-correction)   in  the  type IIA 
    string  partition function on the disk  would appear to  capture particular  quantum $1/N$ terms 
    in the M2 brane  perturbation theory in the   $ 1 \ll k \ll N \ll k^5$ regime 
    \begin{align} \no 
    Z= & {1\ov \sqrt{2\pi}}  e^{2\pi\,T}\,  {\sqrt T \ov \gs } \Big[   1 +  b_1 T^{-1} + b_2 T^{-2} + ...    + O( \gs^2/T) \Big] \\
    = & { k\ov 4 \pi}  e^{ \pi \sqrt{ N/k} } \Big[   1 +  2 \pi  b_1 \sqrt{ k\ov N}  + ( 2\pi)^2 b_2  { k \ov N}  + ...   
     + O( 1/k^2 ) \Big] \la{34}
     \end{align}
   happen   to describe  higher  loop   corrections in  M2  brane   regime  as 
   expressed in terms of  the  M2 brane  parameters.  But  does that  make sense on M2 brane side ... ?
   \fi

One natural generalization of our calculation is to consider the $1\ov6$-BPS Wilson loop 
\ci{Drukker:2008zx,Chen:2008bp}. In this case the localization result derived in \cite{Klemm:2012ii}, expanded in the large $N$ fixed $k$ limit, gives
\be 
\langle W_{\frac{1}{6}}\rangle=   {i\ov  2 \sin{({2 \pi \ov k})} }  \sqrt{{2N\ov k}}\ e^{\pi \sqrt{2 N\ov k} }\big(1+\ldots \big)\ . 
\label{W16}
\ee
Note that there is  an extra factor  $i \sqrt{2N/k}$  compared to $\ha$-BPS case. 
The origin of this factor should be similar to what was discussed in the corresponding string case, where it was argued \cite{Drukker:2008zx} that the string solution should be smeared over a $\CP^1$ in $\CP^3$, leading to two zero modes and hence an overall factor $(\sqrt{T})^2\sim \sqrt{\lambda}$ in the partition function \cite{Drukker:2010nc, Giombi:2020mhz}. For the M2 brane, we similarly expect that the solution relevant to the
 $1\ov6$-BPS  case should be smeared over a $\CP^1$, leading again to an extra tension dependent prefactor $\sim \sqrt{N}$. It would be interesting to study the fluctuation spectrum of the corresponding M2 brane in detail and reproduce from a one-loop calculation the remaining normalization factor in (\ref{W16}).    

Another interesting extension would be to explore the defect CFT defined by the $\ha$-BPS Wilson loop in the large $N$, fixed $k$ limit. The corresponding problem in the type IIA string regime was studied in \cite{Bianchi:2020hsz}. In particular, in that case one finds that the 8+8 fluctuation modes about the AdS$_2$ string solution form a short supermultiplet containing the displacement operator. The same multiplet appears for the M2 brane as the $n=0$ mode in the KK reduction. It would be interesting to understand the interpretation of the higher KK modes from the defect CFT point of view, and compute their boundary correlation functions.\footnote{Similar calculations in the context of the D3 and D5 brane dual
 to the $\ha$-BPS Wilson loop in ${\cal N}=4$ SYM were done in \cite{Giombi:2020amn}, extending the earlier string theory calculation in \cite{Giombi:2017cqn}.} 

\section*{Acknowledgments}
We thank Matteo Beccaria  and Nadav Drukker for useful  comments on the draft, and   also Marcos Marino and Kazumi Okuyama for  correspondence.
 The work of SG is supported in part by the US NSF under Grant No.~PHY-2209997.
AAT is supported by the STFC grant ST/T000791/1.

\small

\bibliographystyle{ssg}
\bibliography{M2-bib}

\end{document}